\documentstyle[11pt,newpasp,twoside]{article}
\def\lapp{\ifmmode\stackrel{<}{_{\sim}}\else$\stackrel{<}{_{\sim}}$\fi}
\def\gapp{\ifmmode\stackrel{>}{_{\sim}}\else$\stackrel{>}{_{\sim}}$\fi}

\def\edcomment#1{\iffalse\marginpar{\raggedright\sl#1\/}\else\relax\fi}
\reversemarginpar

\begin{document}
\title{Neutron Star/Supernova Remnant Associations}
 \author{Victoria M. Kaspi}
\affil{Department of Physics and Center for Space Research, 
Massachusetts Institute of Technology, 70 Vassar Street, Cambridge, MA 02139}

\begin{abstract}
The evidence for associations between neutron stars and supernova
remnants is reviewed.  After summarizing the situation for young radio
pulsars, I consider the evidence from associations that young neutron stars
can have properties very different from those of radio pulsars.  This, though 
still controversial, shakes our simple perception of the Crab pulsar as
prototypical of the young neutron star population.

\end{abstract}

The prediction by \nocite{bz34} Baade \& Zwicky (1934) that neutron
stars are formed in supernova explosions remains one of astronomy's
boldest assertions and greatest triumphs.  Indeed the discovery in 1968 
of the young radio pulsars in the Vela and Crab supernova remnants both
established the neutron-star nature of radio pulsars, and confirmed
the neutron star/supernova remnant connection.  One may well ask why,
over 30 years later, there is still a need to review the subject.

Young neutron stars are interesting for many reasons: they provide one
of the few means of studying the equation of state of nuclear-density
matter, they allow probes of neutron star internal structure from the
study of glitches, they facilitate the study of the magnetospheric
emission mechanism via X-ray and $\gamma$-ray observations, and have
large enough spin-down energies that magnetospheric marvels such as
toroidal relativistic winds and jets can be observed.  On the other 
hand, the study of supernova remnants (SNRs) illuminates the
composition and structure of the progenitors, the physics of core
collapse and of strong shocks, the sites of cosmic ray acceleration,
and the evolution of the interstellar medium.  By associating neutron
stars with SNRs, we obtain information about each class that is
unavailable from either separately.  Associations provide means of
obtaining independent age and distance estimates, which can more
accurately constrain the birth properties of neutron stars, namely,
their period, magnetic field, luminosity and velocity distributions, as
well as help interpret remnant spectra, morphologies, and evolutionary states.

\section{Radio Pulsar/Supernova Remnant Associations}

There are complicated selection effects against finding radio
pulsars and SNRs; both populations are significantly
incomplete, and not in an easily quantifiable way.  However, since the
working hypothesis is that {\it all} pulsars are born in supernovae,
it is fair to consider just the youngest pulsars and ask whether they are
associated with SNRs.

\begin{table}[t]
\caption{Known Rotation-Powered Pulsars having $\tau < 25 $~kyr.}
\begin{tabular}{ccccccc}\tableline
Name  &   $\tau_c$    &  $P$  &    $d$  &    $B$  &  found & SNR\\
      &   kyr    &  s  &   kpc &  $10^{12}$ G &  in   & \\\tableline
B0531+21  &  1.3  &  0.033  &  2.5 & 3.8 &  radio  &   Crab\\
B1509$-$58  &  1.6  &  0.151  &  4.4 & 15  &  X-ray  &  MSH 15$-$5{\it 2}\\
J1119$-$6127&  1.6  &  0.408  &  8.0 & 41  &  radio$^{**}$  & ? \\
B0540$-$69  &  1.7  &  0.050  & 50 & 5.0 &  X-ray  &  N158A\\
J0537$-$6910&  5.0  &  0.016  & 50 & 0.92&  X-ray$^{**}$ &  N157B\\
B1610$-$50  &  7.4  &  0.232  &  7.2 & 11  &  radio$^*$   &  -  \\
J1617$-$5055&  8.0  &  0.069  &  6.5 & 3.1 &  X-ray$^{**}$&  - \\
B0833$-$45 &  11.4  &  0.089  &  0.5 & 3.4 & radio &  Vela\\
B1338$-$62 &  12.1  &  0.193  &  8.7& 7.1&  radio  &  G308.8$-$0.1 \\
B1757$-$24 &  15.5  &  0.124  &  4.6& 4.0&  radio  &  G5.4$-$1.2 \\
B1800$-$21 &  15.8  &  0.134  &  3.9& 4.3&  radio$^*$  & G8.7$-$0.1   \\
B1706$-$44 &  17.4  &  0.102  &  1.8& 3.1&  radio$^*$  &  ? \\
B1853+01 &  20.3  &  0.267  &  2.8& 7.6&  radio$^*$    &  W44\\
B1046$-$58 &  20.4  &  0.124  &  3.0& 3.5&  radio$^*$  &  -\\
B1737$-$30 &  20.7  &  0.607  &  3.3& 17 &  radio$^*$  &  -\\
B1823$-$13 &  21.4  &  0.101  &  4.1& 2.8&  radio$^*$  & ? \\
J1811$-$1926& 24.4  &  0.065  &  5.0& 1.7&  X-ray$^{**}$ & G11.2$-$0.3\\
\tableline
\end{tabular}
\end{table}

Table~1 contains all published radio pulsars having characteristic
ages ($\tau_c \equiv P/2\dot{P}$) under 25~kyr.  This age
cutoff is arbitrary, but the inclusion of only the youngest pulsars is
deliberate: associations involving older pulsars are harder to
evaluate.  First, evidence suggests that SNRs can fade on time scales
of $\sim$20-25~kyr, much shorter than pulsar lifetimes.  Second, a
young pulsar moves a distance $d \simeq 12 (v/450 \; {\rm km/s}) (\tau /
25 \; {\rm kyr})$~pc from its birth place (where $\tau$ is its true age).
This distance can be far enough to reach or escape the parent remnant
shell, depending on the birth velocity distribution, which has been
estimated independently (e.g. \cite{ll94}).  The current version of Green's
SNR catalog\footnote{\tt http://www.mrao.cam.ac.uk/surveys/snrs/} contains 
220 SNRs, 156 of which lie in the range $260^{\circ} < l < 50^{\circ}$
and $|b| < 3.5^{\circ}$.  The area these remnants cover is
$\sim$32 square degrees, or some 3\% of that region of the Galactic
Plane.  In the same area, the published pulsar catalog\footnote{\tt
http://pulsar.princeton.edu/pulsar/catalog.shtml} reports 280 radio
pulsars.  Assuming these are distributed randomly in that area, one
expects $\sim$8.5 chance coincidences.  By contrast, if one restricts
oneself to pulsars having $\tau_c < 25$~kyr, one expects only
$\sim$0.4 chance coincidences.  Thus, very few or none of the
entries in Table 1 should be mere chance superpositions.

In Table~1, the columns are $\tau_c$, period $P$,
distance $d$, inferred surface magnetic field $B$, the discovery band
(items with asterisks were discovered in the past 15 yr, double
asterisks in the past 5 yr), and associated SNR.  Some caveats are
necessary:  $\tau_c$ is only an estimate of the true
age $\tau$; $\tau_c$ is calculated assuming a birth spin period $P_0/P
<< 1$ and braking index $n=3$ and is an overestimate of $\tau$ if $P_0
\simeq P$; the reverse is true for $n < 3$.  Similarly, $B$ is
estimated assuming a dipole braking model, incorrect for objects with
$n<3$, and is dependent on the stellar radius and mass.

Note that of the 4 associations discovered in the past 5~yr, 3 were
found at X-ray energies.  This is in stark contrast to the situation
in the previous 25~yr, in which only 2 of the 13 discovered young pulsars
were found at X-ray energies, the rest being found at radio
wavelengths.  This is due to major advances in X-ray telescopes.
Also, of the 8
pulsars found at radio wavelengths in the past 15~yr, only one was
found by looking for pulsations in a SNR.  This is not for lack of
trying.  Indeed the three major radio pulsar search efforts targeting
SNRs (\cite{kmj+96,gra+96}, Lorimer et al. 1998) \nocite{llc98}
collectively searched 91 targets and found precisely zero young 
pulsars.  By contrast, untargeted surveys of the Galactic plane for
radio pulsars (\cite{jlm+92,cl86,lcm+00}) collectively discovered 6
radio pulsars having $\tau_c < 25$~kyr.  This suggests that the best
way to find young radio pulsars is to look everywhere in the Galactic Plane
{\it but} in SNRs!  One explanation for this quandary is that
bright SNRs reduce radio pulsar survey sensitivity; for example, the
mean radio flux of the SNRs in Green's catalog (omitting the bright
Cas A) roughly doubles the system temperature of the Parkes Multibeam
survey.  The quandary also underscores how incomplete the SNR catalog
is.  Indeed, PSRs B1610$-$50 and J1617$-$5055 ($\tau_c =
7$ and 8~kyr, respectively) appear very young but do not have any
observable associated SNR (Pivovaroff et al. 2000, \cite{kcm+98}).
\nocite{pkg00} The absence of visible SNRs around these
pulsars suggests that remnant fading time can be much shorter than has
been suggested (e.g. Braun et al. 1989). \nocite{bgl89}

\section{Evidence for New Classes of Neutron Stars}

There is now significant evidence that young neutron stars do not all
manifest themselves as radio pulsars.  There has been only a modest
hint that this is true from population statistics.  The supernova rate
in the Galaxy has been estimated to be
$0.025^{+0.008}_{-0.005}$~yr$^{-1}$ (Tammann et al. 1994). \nocite{tls94}
Of these, some 85-90\% are likely to be of Type Ib and II (i.e. producing
compact stellar remnants).  The black hole formation rate is thought to be small, 
at most a few percent (\cite{fry99}).  The pulsar birth rate for radio
luminosities greater than 1~mJy~kpc$^2$ is $0.010 \pm 0.007$~yr$^{-1}$
(\cite{lml+98}).  Given that there must be some low luminosity
pulsars, the agreement in the pulsar birth rates and the
neutron-star-producing supernova rate is reasonable, though a
pulsar dearth is possible.  There has been, however, a long-recognized puzzle 
that most SNRs do not contain visible pulsars.  There are significant
selection effects against finding radio pulsars in SNRs, but this is
less true of finding Crab-like plerions at the centers of shell SNRs, as
those radiate isotropically.  Yet roughly 85\% of Green's catalogued SNRs
have pure shell morphologies.  In fact, independent evidence is mounting
that a significant fraction of young neutron stars have properties
very different from Crab-like radio pulsars; just how large that fraction has
yet to be determined, as is the cause of the diversity.  To summarize,
there are three classes of unusual high-energy sources have been 
compellingly argued to be young, isolated neutron stars:

\smallskip
\noindent
{\bf Anomalous X-Ray Pulsars (AXPs):} The properties of AXPs can be
summarized as follows (see \cite{ms95,gv98}): they exhibit X-ray
pulsations in the range $\sim$5--12~s; they have pulsed X-ray luminosities
in the range $\sim 10^{34}$-$10^{35}$~erg/s; they spin down
regularly within the limited timing observations available (e.g. Kaspi
et al. 1999); \nocite{kcs99} their X-ray luminosities are much greater
than their $\dot{E}$'s; their X-ray spectra are characterized by thermal
emission with $kT \sim 0.4$~keV, with evidence for a hard component;
and they are in the Galactic Plane.  Currently there are 5 confirmed AXPs and
one strong AXP candidate (see Table~2).  Of these 6 sources, 3 lie at the
apparent centers of SNRs: 1E~2259+586 in CTB 109 (\cite{fg81}),
1E~1841$-$045 in Kes 73 (\cite{vg97}), and AX~J1845$-$0258 in
G29.6+0.1 (\cite{gv98,tkk+98}, Gaensler et al. 1999). \nocite{ggv99} 
The association of these
objects with SNRs is arguably the most compelling reason to believe they
are isolated neutron stars.  The leading models explaining their large
X-ray luminosity invoke the large stellar magnetic field as inferred from
the spin down (hence the name ``magnetars''), either using field decay 
(\cite{td96a}) enhanced thermal emission (\cite{hh97}).

\smallskip
\noindent
{\bf Soft Gamma Repeaters (SGRs):} SGRs, of which 4 are known (see
Table~2), are sources that occasionally and suddenly emit bursts of
soft $\gamma$-rays having super-Eddington luminosities.  That 3 of
them lie in the Galactic Plane, and the 4th is in the LMC, 
argues that they are a young population.  The detection of AXP-like X-ray
pulsations from 2 of these sources (e.g. \cite{kds+98}), with evidence
for pulses from the other two, also argues strongly that they are
isolated neutron stars.  Their burst properties and observed spin-down
are well explained in the magnetar model (\cite{td95},
but see Marsden et al. 1999).\nocite{mrl99} The association between SGR
0526$-$66 and the SNR N49 in the LMC (\cite{cdt+82}) first suggested the SGRs
might be young neutron stars, however since then the SGR/SNR
association picture has grown a bit murky.  First, SGR 0526$-$66 is
located near the edge of the N49 shell; this is problematic as it
requires a very high transverse velocity ($v_t >1000$~km/s) for the
SGR (Rothchild et al. 1994).\nocite{rkl94} SGR 1806$-$20 has been
suggested to be associated with the plerionic radio nebula G10.0$-$0.3
(\cite{kf93}), although a recent relocalization of the $\gamma$-ray
source calls the association into question (\cite{hkc+99}).  SGR
1900+14 has been associated with SNR G42.8+0.6
(\cite{vkfg94}), however the $\gamma$-ray source lies well outside the
shell, demanding a distressing $ v_t > 3000$~km/s.  Smith et al. (1999)
\nocite{sbl99} suggest that the newly discovered SGR~1627$-$41 may be
associated with the shell SNR G337.0$-$0.1; the large SGR
positional uncertainty precludes a firm conclusion.

\begin{table}[t]
\caption{Proposed Magnetars and Their Properties.}
\begin{tabular}{cccccccc}\tableline
Type & Name  &  $P$ & $\dot{P}$  &  $\tau_c$ &   B  &  SNR\\
     &       &   s  & $\times 10^{-11}$      &     kyr    & $10^{14}$ G & \\\tableline
AXP & 4U 0142+615   &  8.69  & 0. 23 & 60  &   2.9 & - \\
AXP & 1E 1048$-$5937&  6.45  &  2.2 & 4.6  &   3.8 &  - \\
AXP & RXJ 170849$-$400910& 11.0  & 1.9  & 9.2  & 4.6   &  - \\
AXP & 1E 1841$-$045     & 11.8  & 4.1 & 4.0   & 7.5  & Kes 73 \\
AXP & AX J1845$-$0258    &  6.97  &         &         &         &    G29.6+0.1       \\
AXP & 1E 2259+586       &  6.98  & 0.048 & 210   & 0.59 & CTB 109 \\\tableline
SGR & 0526$-$66 &  8     &          &        &         &       N49 \\
SGR & 1627$-$41 &  6.41? &          &        &         &   G337.0$-$0.1?  \\
SGR & 1806$-$20 &  7.47  &  8.3 & 1.4 &  8   & G10.0$-$0.3? \\
SGR & 1900+14  &  5.16  &  12&  0.68  &  8 & G42.8+0.6?  \\\tableline
\end{tabular}
\end{table}

\smallskip
\noindent
{\bf ``Quiescent'' Neutron Stars:} There are currently 4 cases of
X-ray point sources in SNRs that may be ``quiescent'' neutron stars having
low $\dot{E}$ -- they exhibit neither magnetopsheric
emission nor obvious Crab-like plerions.  These are: 1E~1207.4$-$5209
in PKS ~1209$-$52 (G296.5+10.0,
\cite{hb84,vka+97}), 1E~161348$-$5055 in RCW~103 (\cite{tg80}),
1E~0820$-$4250 in Puppis A (Petre et al. 1996) \nocite{pbw96} although
claimed 75~ms X-ray pulsations, if confirmed, imply that it is an
ordinary rotation-powered pulsar (Pavlov et al. 1999), \nocite{pzt99}
and the newly discovered point source in Cas~A (\cite{tan99}).  That
these sources are only seen in X-rays suggests they could be thermally cooling
neutron stars, still hot following their formation.  Problematic in the neutron 
star interpretation for 1E~161348$-$5055 is that its X-ray luminosity is
apparently variable (Gotthelf et al. 1999).\nocite{gpv99}  In
Cas~A, a preliminary look shows that the point source spectrum may
be harder than expected for a cooling neutron star (M. Pivovaroff,
pers. comm.).  Deep searches for pulsations (enough to see
few percent modulation, as in the known thermally cooling neutron
stars like Vela) and/or high-resolution X-ray spectroscopy to detect
predicted absorption lines in the stellar atmospheres are the most
promising ways of determining the nature of these objects.

\smallskip
In conclusion, although radio pulsars first confirmed the
neutron star/supernova remnant connection, it now appears clear that
they represent only a part of young neutron star phase space.
The origin and full extent of the diversity are not yet clear,
however the problem appears tractable, particularly given the
Parkes Multibeam survey and new and upcoming X-ray missions, including
{\it Chandra}, {\it XMM}, and {\it ASTRO-E}.

\end{document}